\documentclass[pre,showpacs,10pt,amsmath,amssymb]{revtex4}
\usepackage{graphicx}

\begin{document}

\title{Diffusion equations for a Markovian jumping process}

\author
{T. Srokowski and A. Kami\'nska}

\affiliation{
 Institute of Nuclear Physics, Polish Academy of Sciences, PL -- 31-342
Krak\'ow,
Poland }

\date{\today}

\begin{abstract}
We consider a Markovian jumping process which is defined in terms of the
jump-size distribution and the waiting-time distribution with a
position-dependent frequency, in the diffusion limit. We assume the power-law
form for the frequency. For small steps, we derive the Fokker-Planck equation
and show the presence of the normal diffusion, subdiffusion and superdiffusion.
For the L\'evy distribution of the step-size, we construct a fractional
equation, which possesses a variable coefficient, and solve it in the diffusion
limit. Then we calculate fractional moments and define fractional diffusion
coefficient as a natural extension to the cases with the divergent
variance. We also solve the master equation numerically and demonstrate that
there are deviations from the L\'evy stable distribution for large wave numbers.
\end{abstract}


\maketitle

\section{Introduction}

The diffusion is called anomalous if the mean squared displacement of the
Brownian particle does not rise with time linearly, as it is the case for 
the usual Brownian motion, but slower (subdiffusion) or faster
(superdiffusion). There are many examples of physical systems which exhibit
such anomalous transport; they involve complex systems, disordered systems,
semiconductors, polymers, glasses, turbulent plasma, and many others (for a
review see \cite{met,bou}). Physical situations in which the deviation of the linear
dependence is expected comprise: a porous and inhomogeneous environment with traps and
barriers, a coupling to a fractal heat bath via a random matrix interaction
\cite{lut}, long-range and/or long-time correlations. 
Also the transport phenomena in
dynamical hamiltonian systems often exhibit anomalous behaviour because
the corresponding phase space can contain some regular structures which act as dynamical traps
\cite{kar,zas}. A trajectory sticks to such self-similar hierarchy of
islands in chaotic environment and abide on regular paths for a long time. 
The emergence of the anomalous diffusion means that the traditional approach, involving the standard Fokker-Planck equation (FPE) with the diffusion constant, is no longer valid. Usually, the
anomalous diffusion is attributed to memory effects, which are especially
pronounced in non-Markovian descriptions, e.g. by means of the generalized
Langevin equation \cite{kubo} and the continuous-time random walk models (CTRW)
\cite{met}. In the diffusive limit, those approaches resolve themselves 
to the fractional equations formalism \cite{met,bar1,sca} which
contain the Riemann-Liouville operator as the fractional time-derivative; the step-size distribution is the Gaussian.
As a direct consequence of the nonlocality in time and, therefore, 
of the lack of Markovian property, the uncoupled CTRW predicts subdiffusive
solutions in this case \cite{kla1,bar}. However, the anomalous behaviour
is possible also for the Markovian processes; they are described by FPE with the position-dependent diffusion term
and physically can correspond e.g. to the diffusion on
fractal objects \cite{osh} or to the Langevin equation
with the multiplicative noise \cite{kwo}. In this paper we discuss
another example of such Markovian process.  

A more general approach considers the L\'evy distribution which is
defined, in its symmetric version, via the characteristic function
$\phi(k)=\exp(-\alpha k^\mu)$, where $0<\mu\le2$. If $\mu\ne2$ the
second moment is infinite and the generalized central limit theorem must
be applied. Physically, L\'evy processes can reflect self-similarity:
they are the stable solutions of the renormalization group method and
are invariant under the scaling of position and time. In contrast to the
Gaussian distribution, they include large fluctuations. The L\'evy
distributions are present in many problems connected with various
branches of since, including not only physics but also biology, economy,
financial market research, etc. The diffusion equation for the L\'evy
process involves a fractional derivative over the process value 
\cite{com,met1,met2,zas} and for $\mu=2$ it resolves itself to the
ordinary FPE. Since the variance diverges for $\mu<2$ and 
the traditional description of the diffusion process is no longer valid,
one has to introduce new concepts, e.g. to study fractional moments 
\cite{com1} or to restrict the integration in averaging to 
a finite box which scales with time in a prescribed way \cite{fog,com2}.

Problems related to generalized diffusion equations, which contain either anomalous behaviour of the variance or infinite fluctuations, are the subject of the present paper. We deal with a jumping process which is Markovian, defined in terms of a jump-size distribution  $Q(x)$ and the waiting time distribution $P_P(\tau)$. A peculiarity of the distribution $P_P(\tau)$ consists in a position-dependent jumping rate. The process is defined in Sec.II. In Sec.III we derive the FPE as a small step limit of the master equation. Sec.IV is devoted to the fractional diffusion equation which is an approximation of the master equation in the diffusion limit for $Q$ given by the L\'evy distribution; we solve this equation, discuss its diffusion properties and compare with numerical solution of the master equation. The results of our analysis are discussed in Sec.V.

\section{Definition of the process}

The jumping process we are to deal with in this paper is a stepwise constant stochastic process $x(t)$ defined in terms of two probability distributions \cite{kam1}. 
The waiting time density distribution determines
the time intervals $\tau_i$ between consecutive jumps and it is assumed in the
Poissonian form:
\begin{equation}
\label{poi}
P_P(\tau)=\nu(x){\mbox e}^{-\nu(x)\tau},
\end{equation}
where the jumping rate $\nu(x)$ depends on the process value (the position). 
The size of the jumps is determined by a given normalized distribution $Q(r=x-x')$. 
Then the infinitesimal transition probability can be easily constructed and 
the master equation derived. It is of the form
  \begin{equation}
  \label{fpkp}
  \frac{\partial}{\partial t}p(x,t) = -\nu(x)p(x,t) +
  \int Q(x-x')\nu (x') p(x',t) dx'.
  \end{equation}

Because of the variable jumping rate $\nu(x)$, the above process can be
regarded as a version the kangaroo process \cite{fri,kam}. The
difference consists in the jump-size dependence of $Q$ -- in the kangaroo process $Q$ depends only on the current position. Due to that property, the Eq.(\ref{fpkp}) can describe transport phenomena \cite{kam1}.
On the other hand, the process constitutes a special case of the coupled
CTRW. Taking into account the position dependence of jumping rate is an
important generalization of traditional random walk approaches. 
Such dependence is expected in many phenomena in which inhomogeneity of the environment cannot be neglected \cite{san}.

The stationary solution of Eq.(\ref{fpkp}) can be
easily obtained: $P(x)=1/\nu(x)$. The normalization condition imposes
restrictions on the function $\nu(x)$: it must rise sufficiently fast in the
infinity and sufficiently slow near zero. If, in turn, $1/\nu(x)$ has
poles at some $x$, the stationary solution also exists and it is in the
form of a combination of the delta functions. 
In the other cases $P(x)$ does not exist. 
A special version of the stationary process, defined on the circle, 
exhibits long-time correlations and can serve as a model of the $1/f$ 
noise \cite{sro}.

The general, time-dependent solution of the Eq.(\ref{fpkp}) can be obtained by
using the Laplace transforms. The formal expression for the Laplace transform
of $p(x,t)$ is the following \cite{kam1}
  \begin{widetext}
  \begin{equation} 
  \label{pwarl}
  \begin{split}
  &{\widetilde p}(x,s)=\frac{p_0(x)}{s+\nu(x)}+\frac{1}{s+\nu(x)}
  \int\frac{\nu(x_0)p_0(x_0)Q(x-x_0)}{s+\nu(x_0)}dx_0+ \\
  & +\frac{1}{s+\nu(x)} \sum_{k=2}^{\infty}
  \int\frac{\nu(x_0)p_0(x_0)}{s+\nu(x_0)} Q(x-x_{k-1})
  \left[\prod_{i=2}^k\frac{\nu(x_{i-1})Q(x_{i-1}-x_{i-2})}
  {s+\nu(x_{i-1})}dx_{i-1}\right]dx_0,
  \end{split}
  \end{equation}
  \end{widetext}
where $p_0(x)$ stands for the initial condition.

By multiplying the Eq.(\ref{fpkp}) by $x^2$ and by integrating over $x$, one can obtain the equation which governs the time evolution of the variance.
Assuming that $Q$ has the vanishing mean and finite second moment, we yield
the following result:
\begin{equation}
\label{var}
\frac{\partial\langle x^2\rangle_p}{\partial t}=
\langle x^2\rangle_Q\langle\nu\rangle_p.
\end{equation}
The simple case $\nu(x)=$const can be solved completely and the Eq.(\ref{var})
predicts the normal diffusion. However, if $\langle x\rangle_Q$ does not vanish,
the diffusion becomes ballistic \cite{kam1}. For $Q$ with divergent second
moment, in turn, $\langle x^2\rangle_p$ is infinite.

\section{Small jumps: the Fokker-Planck equation}

The expression (\ref{pwarl}) is difficult to handle and one has to resort to approximations. 
In the limit of small jumps, the process becomes continuous both in space and
time and FPE may be a candidate for such approximation. In order to construct it, we apply the Kramers-Moyal expansion \cite{gar} of the master
equation. In that method, one changes the integration variable 
in the Eq.(\ref{fpkp}) by introducing the step size $r=x-x'$ and,
after the expansion of the function $p(x-r,t) \nu(x-r)$ around $r=x$, 
one obtains the master equation in a form of the following series
\begin{equation} 
\label{szereg}
\frac{\partial p(x,t)}{\partial t} = \sum_{n=1}^{\infty}
\frac{(-1)^n}{n!} \langle r^n \rangle_Q (\frac{\partial}{\partial
 x})^n [p(x,t) \cdot \nu(x)]
\end{equation}
which is still exact. The approximation consists in neglecting all terms of
the order higher than two. Obviously, all moments of $Q$ must be finite. 
Finally, we obtain the FPE:
\begin{equation} 
\label{rfp}
   \frac{\partial p(x,t)}{\partial t}=\frac{\sigma^2}{2}
   \frac{\partial^2 [\nu(x)p(x,t)]}{\partial x^2},
   \end{equation}
where $\sigma^2=\langle r^2\rangle_Q$. Therefore, the jumping rate may be
interpreted as the position-dependent diffusion coefficient. The approximation
is valid if the jumps 
are small and the function $\nu(x)$ is smooth \cite{vkam}:
\begin{equation}
\label{mac}
\begin{array}{cll}
Q(r)\approx0 & {\rm for} & r>\delta\\
 \nu(x+\Delta x)\approx \nu(x) &{\rm for} & \Delta x<\delta,
\end{array}
\end{equation}
where $\delta$ is a small parameter. 

For $\nu(x)=$const, Eq.(\ref{rfp}) takes the form of the ordinary diffusion
equation which describes the Wiener process and is characterized by the normal
diffusion. An interesting case is the power-law form of the jumping
frequency: 
\begin{equation} 
\label{nux}
\nu(x)=\gamma |x|^{-\theta},
\end{equation}
where $\theta$ is a constant parameter and $\gamma$ ensures proper units; in the
following we assume $\gamma=1$. The diffusion coefficient in the form
(\ref{nux}) has been used to describe the diffusion on fractal objects
\cite{osh}, the transport of fast electrons in a hot plasma \cite{ved},
turbulent two-particle diffusion \cite{fuj}. The FPE with that diffusion
coefficient has been analyzed from the point of view of the first passage time
in Ref. \cite{kwo1}. The FPE (\ref{rfp}) with $\nu$ given by Eq.(\ref{nux}) can
be solved exactly \cite{hen}: 
   \begin{equation}
   \label{rozw}
   p(x,t)=C_{\theta} \frac{|x|^{\theta}\exp(-\frac{2|x|^{2+\theta}}
{\sigma^2 (2+\theta)^2t})}{(\sigma^2t/2)^{\frac{1+\theta}{2+\theta}}},
   \end{equation}
where $C_{\theta}=\frac{1}{2 \Gamma(\frac{1+\theta}{2+\theta})}
|2+\theta|^{\frac{\theta}{2+\theta}}$. Some values of $\theta$ must 
be excluded. Since for $\theta<-3$ the 
Eq.(\ref{fpkp}) has the stationary solution (in the form of the delta
function), the approximation by the FPE is not valid; the diffusion in this
case is local. Moreover, for $\theta\in (-2,-1)$ the distribution (\ref{rozw})
is not normalized. Therefore, we have either $\theta\in (-3,-2)$ or 
$\theta>-1$. 
\begin{figure}[htb]
\includegraphics[width=8.5cm]{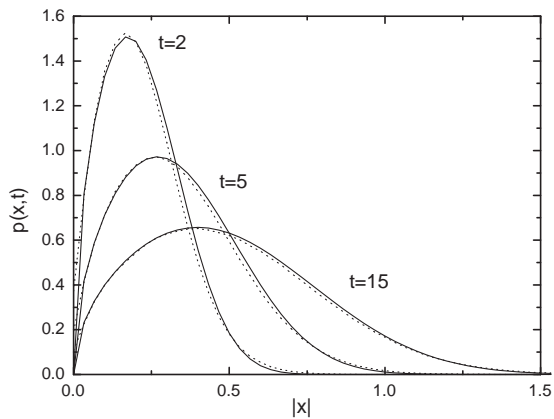}
\caption{Time evolution of the FPE solutions $p(x,t)$ calculated from
the Eq.(\ref{rozw}) (solid line) and solutions of the master equation
(\ref{fpkp}) for the Gaussian form of the distribution $Q$ (dots). Parameters
are: $\theta=0.5$ and $\sigma=0.1$.}
\end{figure}

Fig.1 presents the comparison of the FPE solutions $p(x,t)$, calculated according to the Eq.(\ref{rozw}), with the solutions of the master equation (\ref{fpkp}), obtained from Monte Carlo simulations of the stochastic trajectories. The agreement is good already for $\sigma=0.1$.

The diffusion properties of the system follow directly from the solution
(\ref{rozw}). The mean squared displacement for $\theta>-1$ is given by the
power-law function of time by the following formula
   \begin{equation}
   \label{w}
   \langle x^2(t) \rangle=
   \frac{\Gamma(\frac{3+\theta}{2+\theta})}{
   \Gamma(\frac{1+\theta}{2+\theta})}
   \left[\frac{\sigma^2}{2}(2+\theta)^2t\right]^{\frac{2}{2+\theta}}
   \end{equation}
and the diffusion coefficient is:
\begin{equation}
\label{dif2}
{\cal D}=\frac{\langle x^2\rangle}{2t}\sim t^{-\theta/(\mu+\theta)}~~~~~~(t\to\infty).
\end{equation}
Then we can distinguish three cases. For $\theta\in(-1,0)$, ${\cal D}$ is infinite 
and the superdiffusion emerges. For $\theta>0$, ${\cal D}=0$ and we
get the subdiffusion. Finally, the normal diffusion takes place for
$\theta=0$. Therefore, the jumping process involves all kinds of diffusion.

\section{The fractional diffusion equation}

Let us now assume the distribution $Q$ in the form of the symmetric L\'evy distribution, defined by its Fourier transform:
\begin{equation}
\label{lev}
{\widetilde Q}(k)=\exp(-K^\mu |k|^\mu)~~~~~~~~~~(K>0),
\end{equation}
where $1<\mu<2$. In contrast to the Gaussian distribution, $Q(x)$, corresponding to the Eq.(\ref{lev}), has algebraic tails, $Q(x)\sim |x|^{-1-\mu}~(|x|\to\infty)$, which makes long jumps very probable. In the diffusion limit $k\to0$, the Eq.(\ref{lev}) is given by ${\widetilde Q}(k)\approx 1-K^\mu |k|^\mu$. We wish to derive an equation which could serve as an approximation to the master equation (\ref{fpkp}) in the diffusion limit. We take the Fourier transform of the Eq.(\ref{fpkp}) and insert ${\widetilde Q}(k)$ in the above form, that yields
\begin{equation}
\label{fracek}
\frac{\partial{\widetilde p}(k,t)}{\partial t}=-K^\mu |k|^\mu{\cal F}[\nu(x)p(x,t)].
\end{equation}
The Eq.(\ref{fracek}) can be formally inverted by expressing the invert Fourier transform by a fractional Weyl-Riesz operator: ${\cal F}^{-1}(-|k|^\mu)=\frac{\partial^\mu}{\partial|x|^\mu}$ \cite{old}. The resulting equation is the following
\begin{equation}
\label{frace}
\frac{\partial p(x,t)}{\partial t}=K^\mu\frac{\partial^\mu[\nu(x)p(x,t)]}{\partial|x|^\mu}.
\end{equation}
Our aim is to solve the Eq.(\ref{frace}) for $\nu$ given by Eq.(\ref{nux}), where $\theta>-1$, with the initial condition $p(x,0)=\delta(x)$. Obviously, the presence of the $x$-dependent term under the fractional derivative poses the main difficulty.

\subsection{The case $\theta=0$}

The Eq.(\ref{fracek}) for this case can be easily solved:
\begin{equation}
\label{solk0}
{\widetilde p}(k,t)=\exp(-K^\mu t|k|^\mu).
\end{equation}
The above expression is the characteristic function of the L\'evy process \cite{zas} and its inversion produces the symmetric L\'evy distribution. To handle those distributions, it is convenient to apply Fox functions. In fact, the L\'evy distribution in its most general form can be expressed as a Fox function \cite{sch}. That formalism is  useful in describing stable processes because it reflects scaling properties of the underlying phenomena. The definition and some properties of the Fox functions are presented in Appendix. 

The solution of the Eq.(\ref{frace}) can then be expressed in the following way \cite{wes,met}
\begin{eqnarray} 
\label{solp0}
p(x,t)=\frac{1}{\mu|x|}H_{2,2}^{1,1}\left[\frac{|x|}{(K^\mu t)^{1/\mu}}\left|\begin{array}{l}
(1,1/\mu),(1,1/2)\\
\\
(1,1),(1,1/2)
\end{array}\right.\right].
\end{eqnarray}
The asymptotics $|x|\to\infty$ results from the expansion (\ref{A.3}): $p(x,t)\sim t/|x|^{1+\mu}$ and implies that the variance, as well as all fractional moments of the order of $\mu$ or higher, diverge.

\subsection{The general case}

The Eq.(\ref{frace}), as an approximation to the master equation in the diffusion limit, has been obtained from the Eq.(\ref{fpkp}) by dropping all terms of the order higher than $k^\mu$ in the expansion of $Q(k)$. Consequently, solving the Eq.(\ref{frace}), we can neglect those terms without introducing any additional idealization.  We look for a solution in the form of the Fox function
\begin{eqnarray} 
\label{s1}
p(x,t)=NaH_{2,2}^{1,1}\left[a|x|\left|\begin{array}{c}
(a_1,A_1),(a_2,A_2)\\
\\
(b_1,B_1),(b_2,B_2)
\end{array}\right.\right],
\end{eqnarray}
similar to that for the case $\theta=0$, where $a=a(t)$ and $N$ is the normalization constant. We want to determine the coefficients $a_i$, $A_i$, $b_i$ and $B_i$, as well as the function $a(t)$, by expansion of both sides of the Eq.(\ref{fracek}) in consecutive fractional powers of $k$ and neglect the terms which are small for $k\to0$. 

The function $p_\theta=x^{-\theta}p(x,t)$ on the right hand side of Eq.(\ref{fracek}) can be expressed as the Fox function due to the multiplication rule (\ref{A.5}). Its Fourier transform, given by the Eq.(\ref{A.6}), expanded according to the formula (\ref{A.3}), takes the form
\begin{equation}
\label{rhs}
{\widetilde p_\theta}(k,t)=2\pi(h_0^{(\theta)}+h_1^{(\theta)}k'+O(k^\mu)),
\end{equation}
where $k'=Kk/a$. Deriving explicitly consecutive terms and utilizing simple properties of the gamma function, one can find conditions for the coefficients. First, to get the term of order $k^0$, we introduced the condition $1-a_1=A_1(1-\theta)$. $h_1^{(\theta)}=0$ identically because the gamma function in the denominator is infinite. Similarly, for the function $p(x,t)$ we have:
\begin{equation}
\label{lhs}
{\widetilde p}(k,t)=2\pi(h_0+h_{-\theta}k'^{-\theta}+h_1k'+h_2k'^2+h_\mu k'^\mu+O(k^{2\mu+\theta})),
\end{equation}
where we imposed the condition $2-a_1=A_1(1+\mu)$ to get the exponent
$\mu$ for the term $k^\mu$. Then the above conditions determine the
first two coefficients of the Fox function: $a_1$ and $A_1$. We need also
$h_{-\theta}=0$; this requirement can be satisfied if the argument of one of the gamma functions in the denominator assumes an integer and non-positive value. The condition for that is $1-b_2-B_2(1-\theta)=0$
(alternatively, we could impose a similar condition for $a_2$ and
$A_2$). The same procedure allows us to satisfy the requirement $h_2=0$ and to determine
$B_2=1/(2+\theta)$. Finally, since the coefficients $a_2$ and $A_2$
enter the above expressions in a similar way as $b_2$ and $B_2$, we want
to preserve this symmetry, present for the case $\theta=0$, and assume
$a_2=b_2$ and $A_2=B_2$. 

Then we insert the expansions Eqs.(\ref{rhs}) and (\ref{lhs}) into the Eq.(\ref{fracek}) and separate the time-dependent term: $(a/K)^{-\mu-\theta-1}\dot a/K=-\lambda$, where $\lambda$ is a constant which scales the time and then it is not essential. We assume $\lambda=1/(\mu+\theta)$. By taking into account the initial condition for the Eq.(\ref{frace}), one can write down the solution for the function $a(t)$ as:
\begin{equation}
\label{aodt}
a=Kt^{-\frac{1}{\mu+\theta}}
\end{equation}
and then reduce the problem to a simple equation: 
\begin{equation}
\label{hmt}
-\frac{\mu}{\mu+\theta}h_\mu=h^{(\theta)'}_0,
\end{equation}
where $h^{(\theta)'}_0=a^{-\theta}h^{(\theta)}_0$.

After implementing the coefficients we have evaluated, we obtain for the term
$h_\mu$ the following expression:
$h_\mu=\pi^{-2}(\mu+\theta)\Gamma(-\mu)\Gamma(b_1+B_1(1+\mu))\cos(\mu\pi/2)\sin(
\frac{\mu+\theta}{2+\theta}\pi)$. Unfortunately, the term $h^{(\theta)'}_0$
cannot be obtained directly from the series expansion because the undetermined
expression emerges. Instead, we proceed as follows. $h^{(\theta)'}_0$ can be
expressed as: 
$h^{(\theta)'}_0=(2\pi)^{-1}a^{-\theta}\widetilde{p}_\theta(k=0)=\pi^{-1}\int_0
^\infty z^{-1}W(z)dz$, where we obtained the function
\begin{eqnarray} 
\label{wau}
W(z)=H_{2,2}^{1,1}\left[z\left|\begin{array}{l}
(1,\frac{1}{\mu+\theta}),(1,\frac{1}{2+\theta})\\
\\
(b_1+B_1(1-\theta),B_1),(1,\frac{1}{2+\theta})
\end{array}\right.\right]
  \end{eqnarray}
by applying the relation (\ref{A.5}). The required term can now be easily evaluated as the Mellin transform ${\cal M}(W)(s)=\int_0^\infty W(z)z^{s-1}dz$: 
\begin{equation}
\label{h0t}
h^{(\theta)'}_0=\pi^{-1}{\cal
M}(W)(s=0)=\pi^{-1}\chi(0)=\pi^{-1}\lim_{\epsilon\to
0}\chi(\epsilon)=\frac{1}{\pi}\frac{\mu+\theta}{2+\theta}\Gamma(b_1+B_1(1-\theta
)).
\end{equation}

Inserting the expressions for $h_\mu$ and $h^{(\theta)'}_0$ to the
Eq.(\ref{hmt}), yields the relation between $b_1$ and $B_1$. Then, finally, the
solution of the Eq.(\ref{frace}) reads
\begin{eqnarray} 
\label{solp}
p(x,t)=NaH_{2,2}^{1,1}\left[a|x|\left|\begin{array}{l}
(1-\frac{1-\theta}{\mu+\theta},\frac{1}{\mu+\theta}),(1-\frac{1-\theta}{2+\theta},\frac{1}{2+\theta})\\
\\
(b_1,B_1),(1-\frac{1-\theta}{2+\theta},\frac{1}{2+\theta})
\end{array}\right.\right]
  \end{eqnarray}
with the condition
\begin{equation}
\label{b1b1}
\frac{\mu}{\pi}\frac{2+\theta}{\mu+\theta}\Gamma(-\mu)\Gamma(b_1+B_1(1+\mu))
\sin\left(\frac{\mu+\theta}{2+\theta}\pi\right)\cos(\mu\pi/2)+\Gamma(b_1+B_1(1-\theta))=0
\end{equation}
which follows directly from the Eq.(\ref{hmt}). The general theory of the Fox functions implies that $p(x,t)$ is the analytic function for all $x\ne0$ if $B_1>1/(\mu+\theta)$. Apart from that --very weak-- condition, one parameter is arbitrary. We will return to this issue in Subsec. D.
The normalization factor $N$ can be determined in a simple way by using the Mellin transform: 
$N=[2\chi(-1)]^{-1}$, that yields
\begin{equation} 
\label{nor}
N=-\frac{\pi}{2}\left[\Gamma(b_1+B_1)\Gamma\left(-\frac{\theta}{\mu+\theta}
\right)\sin\left(\frac{\theta}{2+\theta}\pi\right)\right]^{-1}.
\end{equation}
In the $k$-space, the solution (\ref{solp}) is of the form
\begin{equation}
\label{plim}
{\widetilde p}(k,t)= 1-\sigma^\mu k^\mu+\dots\approx\exp(-\sigma^\mu k^\mu)
\end{equation}
where
\begin{equation}
\label{sigmu}
\sigma^\mu=\frac{K^{-\mu}}{\mu}\frac{(\mu+\theta)^2}{2+\theta}\frac{\Gamma(b_1+B_1(1-\theta))
\Gamma(-\frac{\theta}{2+\theta})}{\Gamma(b_1+B_1)\Gamma(-\frac{\theta}{\mu+\theta})}\,
t^{\frac{\mu}{\mu+\theta}}
\end{equation}
and we neglected the terms of the order $k^{2\mu+\theta}$ ($\theta<0$) or 
$k^{2\mu}$ $(\theta>0)$.
The formula (\ref{plim}) means that within the scope of validity of our approximation the solution (\ref{solp}) is equivalent to the L\'evy stable distribution.

The Fox functions can be evaluated from series expansions. Since they are poorly convergent, one needs the series both for small and large values of the argument. According to Eq.(\ref{A.3}), the expansion of $p(x,t)$ for small $|x|$ is of the form
\begin{equation}
\label{serm}
p(x,t)=\frac{Na^{1+b_1/B_1}}{\pi B_1}|x|^{b_1/B_1}\sum_{i=0}^\infty\Gamma\left(\frac{1-\theta}{\mu+\theta}+\frac{1}{\mu+\theta}\frac{b_1+i}{B_1}\right)\sin\left[\left(\frac{1-\theta}{2+\theta}+\frac{1}{2+\theta}\frac{b_1+i}{B_1}\right)\pi\right](-1)^i(a|x|)^{i/B_1}/i!.
\end{equation}
Using the property (\ref{A.4}), we obtain the series for large $|x|$:
\begin{equation}
\label{serd}
p(x,t)=-\frac{Na}{\pi}(\mu+\theta)\sum_{i=1}^\infty\Gamma[b_1+B_1(1-\theta+i(\mu+\theta))]
\sin\left(\frac{\mu+\theta}{2+\theta}i\pi\right)(-1)^i(a|x|)^{-1+\theta-i(\mu+\theta)}/i!.
\end{equation}
The above expression implies that the tail of the distribution has the same $x$-dependence as for the case $\theta=0$: $p(x,t)\sim t^{\mu/(\mu+\theta)}/|x|^{1+\mu}$ ($|x|\to\infty$).

\subsection{Diffusion}

A characteristic feature of the L\'evy distributions is the divergence of moments. In particular, the mean squared displacement is infinite for any time and then the transport phenomena require a modified formalism for the diffusion. One possibility is to substitute the variance by a moment of the order $\delta<\mu$. 

In order to evaluate the moments of the distribution (\ref{solp}), we utilize properties of the Mellin transforms. A simple calculation yields:
\begin{equation}
\label{mom}
\langle |x|^\delta\rangle=2N\int_0^\infty x^\delta p(x,t)dx=2Na^{-\delta}\chi(-\delta-1)=-\frac{2NK^{-\mu}}{\pi}\Gamma(b_1+B_1(1+\delta))\Gamma\left(-\frac{\theta+\delta}{\mu+\theta}\right)\sin\left(\frac{\theta+\delta}{2+\theta}\pi\right)t^{\delta/(\mu+\theta)}.
\end{equation}
Applying the above expression, one can compare individual cases in respect to the speed of transport.  However, as long as the parameter $\delta$ is arbitrary, such formalism seems to be incomplete. Can it be fixed in some way? Clearly, the value $\delta=\mu$ is distinguished. Since that moment is divergent, we consider $\delta=\mu-\epsilon$, where $0<\epsilon\ll\theta$ and then the case $\theta=0$ is excluded. In the limit $\epsilon\to0$, the gamma function can be expanded and Eq.(\ref{mom}) takes the form
\begin{equation}
\label{mom1}
\langle |x|^{\mu-\epsilon}\rangle\approx \frac{2NK^{-\mu}}{\pi\epsilon}(\mu+\theta)\Gamma(b_1+B_1(1+\mu))\sin\left(\frac{\mu+\theta}{2+\theta}\pi\right)t^{\mu/(\mu+\theta)}.
\end{equation}
Let us now define the fractional diffusion coefficient ${\cal D}^{(\mu)}(t)$:
\begin{equation}
\label{dif}
{\cal D}^{(\mu)}\equiv\frac{1}{\Gamma(1+\mu)}\frac{1}{t}\,\lim_{\epsilon\to0^+}\epsilon\langle |x|^{\mu-\epsilon}\rangle~~~~~~(t\to\infty),
\end{equation}
where $1<\mu<2$.
According to (\ref{mom1}), the limit is finite and ${\cal D}^{(\mu)}\sim t^{-\theta/(\mu+\theta)}$. 

The interpretation of the above result is straightforward. 
If $\theta<0$ the coefficient ${\cal D}^{(\mu)}$ rises with time to infinity and we have the"superdiffusion". Conversely, for $\theta>0$ there is the "subdiffusion". Therefore we have obtained formally the same result as for FPE, Eq.(\ref{dif2}), except the variance has been substituted by the fractional moment. The common conclusion, drawn both from the Gaussian and the fractional case, is that the sign of $\theta$ decides which kind of diffusion the system reveals.

\subsection{Numerical examples}

In this Subsection we evaluate the probability distributions for specific cases. They are compared with the solutions of the master equation (\ref{fpkp}), obtained by the Monte Carlo simulations of stochastic trajectories of the jumping process. The L\'evy-distributed jump-size density has been generated by using the algorithm from Ref.\cite{cha}.

It follows from the expansion (\ref{serm}) that $p(x,t)\sim |x|^{b_1/B_1}$ for small $|x|$. Therefore, the ratio of the coefficients $b_1$ and $B_1$ determines the shape of the distribution $p(x,t)$ there. Since our approximation is suited for large $|x|$,  this ratio remains undetermined. The analysis of the master equation indicates that its solutions exhibit the power law dependence of the form $|x|^\theta$ for small $|x|$. We utilize this property and assume the relation $b_1=\theta B_1$ in the present Subsection; $B_1$ follows from the numerical solving of the Eq.(\ref{b1b1}). 
The Fox functions have been evaluated from the series (\ref{serm})
and (\ref{serd}) with sufficient precision for all $|x|\in(0,\infty)$.
\begin{figure}[h]
\includegraphics[width=0.6\textwidth]{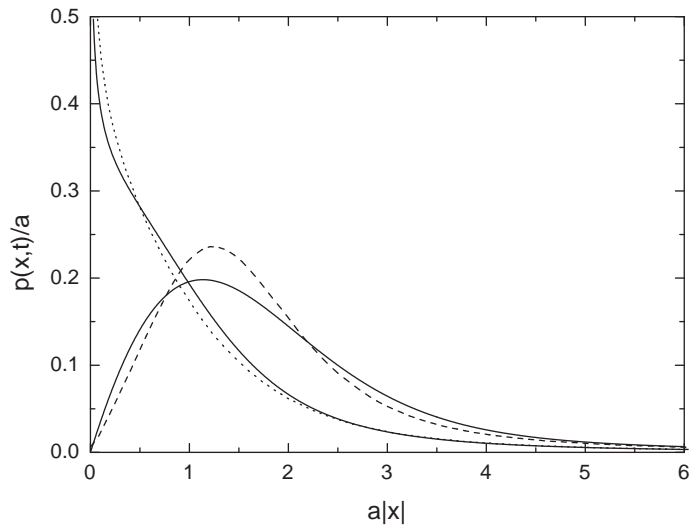}
\caption{The solutions of Eq.(\ref{fpkp}) for $\theta=-0.2$ (dots) 
and $\theta=1$ (dashed line). The corresponding solutions 
(\ref{solp}) are marked by solid lines.}
\end{figure}

The solutions of the fractional equation (\ref{solp}), for
a negative and a positive value of $\theta$, are presented in Fig.2. 
They correspond to the superdiffusion and subdiffusion, respectively,
and display very different shapes. All the distributions with $\theta>0$
rise at small $|x|$ and display a maximum, whereas those for $\theta<0$ 
fall monotonically. 
The comparison with the solutions of the master equation is also
presented in Fig.2. The curves which represent both equations have 
similar shapes and their tails coincide.

Fig.3 presents the comparison of the Fourier transforms for the solution of
both equations; the same cases as in Fig.2 are shown: $\theta=-0.2$ and 
$\theta=1$. For that purpose, the master equation has been solved 
for all $x$ up to very large values, then the
numerical integration with the cosine function has been performed. 
In the case of the master equation solutions,
the figure reveals substantial deviations from the
straight lines -- which represent the shape of the L\'evy distribution --
for large $k$: for $\theta<0$ that distribution  stabilizes with $k$, 
whereas it falls rapidly to zero for $\theta>0$. At small $k$ both solutions 
coincide with those of the fractional equation.

\begin{figure}[h]
\includegraphics[width=0.6\textwidth]{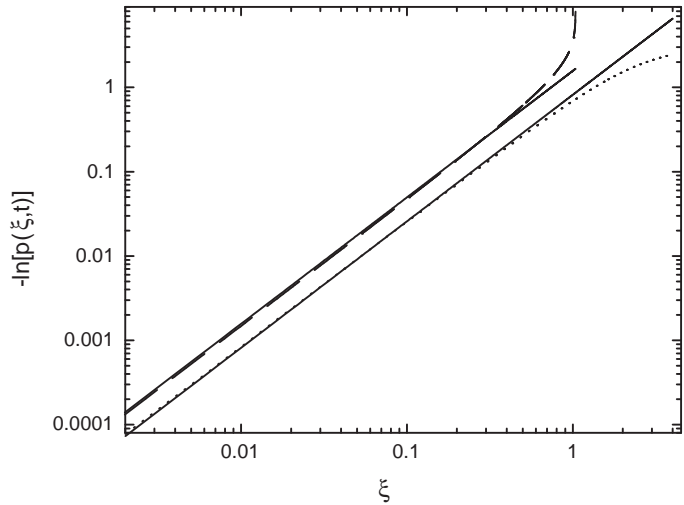}
\caption{Logarithm of the Fourier transform from the solution of 
Eq.(\ref{fpkp}) for $\theta=-0.2$ (dots) and $\theta=1$ (dashed line). 
The same quantity for the solution (\ref{solp}) which corresponds to
both cases, calculated according to
Eq.(\ref{plim}), is marked by solid lines; $\xi=k/a$.}
\end{figure}

\section{Discussion}

The jumping process we have discussed in this paper is Markovian because the waiting time probability distribution has the exponential form. However, since the jumping frequency depends on the process value, the system possesses some properties which are usually attributed to non-Markovian processes. In particular, we have demonstrated the presence of the anomalous diffusion. 

If the step-size is small, the jumping process can be regarded as continuous and described in terms of the FPE.
That approximation has been accomplished by applying the Kramers-Moyal expansion, on the assumption that all moments of the step-size distribution are finite. It has been demonstrated --by solving the FPE exactly for the power-law frequency $\nu(x)$-- that both normal and anomalous diffusion emerges.

On the other hand, we have considered the diffusion limit of the master equation 
for the step-size distribution which is stable and has divergent moments. The resulting equation (\ref{frace}) is fractional and contains a variable coefficient, in contrast to usually studied equations which govern L\'evy processes. We have demonstrated that in the diffusion limit the Eq.(\ref{frace}) is satisfied by the Fox function if $\nu$ depends algebraically on the position. The coefficients of the Fox function have been derived by inserting it to the equation and by comparing the terms. However, one parameter must remain undetermined if only  the diffusion limit is considered because it is responsible for the behaviour of the solution at small $|x|$. 

The solution, since it is expressed as the Fox function, depends on time in the scaling form. 
Due to that property, simple conclusions about the transport can be drawn. The fractional moments are given by complicated expressions but the time-dependence factorizes and it obeys a simple power law. We have defined the fractional diffusion coefficient ${\cal D}^{(\mu)}$ which allows us to establish a correspondence with the standard description in terms of the FPE.  Both approaches predict the diffusion coefficient, either ${\cal D}$ or ${\cal D}^{(\mu)}$, in the form $\sim t^{-\theta/(\mu+\theta)}$, i.e. the subdiffusion ($\theta>0$) and the superdiffusion ($\theta<0$). We would like to emphasize that the above conclusions are independent of a specific choice of the free parameter in the solution.

An independent numerical analysis of the master equation (\ref{fpkp}) 
provides additional information about the jumping process. 
We can learn that it is not the L\'evy stable process: substantial deviations 
from the L\'evy distribution at large $k$ are clearly visible. They 
become meaningless in the diffusion limit.

The comparison of the solutions of both diffusion equations we have
discussed in this paper with the results of the numerical analysis of
the master equation shows a reasonable agreement not only in the
asymptotic limit of large $|x|$. Since the expressions (\ref{rozw}) and
(\ref{solp}) are relatively simple, compared to the Eq.(\ref{pwarl}), both
diffusion equations could serve as convenient approximations to the
master equation (\ref{fpkp}). 

\section*{APPENDIX}

\setcounter{equation}{0}
\renewcommand{\theequation}{A\arabic{equation}}

The Fox function \cite{fox,mat,sri} (for a review of the most important properties see \cite{sch,met1}) is defined as an inverse Mellin transform in the following way:
  \begin{eqnarray} \label{A.1}
H_{pq}^{mn}\left[z\left|\begin{array}{c}
(a_p,A_p)\\
\\
(b_q,B_q)
\end{array}\right.\right]=H_{pq}^{mn}\left[z\left|\begin{array}{c}
(a_1,A_1),(a_2,A_2),\dots,(a_p,A_p)\\
\\
(b_1,B_1),(b_2,B_2),\dots,(b_q,B_q)
\end{array}\right.\right]=\frac{1}{2\pi i}\int_L\chi(s)z^sds,
  \end{eqnarray}
where 
\begin{equation}
\label{A.2}
\chi(s)=\frac{\prod_1^m\Gamma(b_j-B_js)\prod_1^n\Gamma(1-a_j+A_js)}
{\prod_{m+1}^q\Gamma(1-b_j+B_js)\prod_{n+1}^p\Gamma(a_j-A_js)}.
\end{equation}
Coefficients $A_i$ and $B_i$ are positive. The contour $L$ is a straight line parallel to the imaginary axis which separates the poles of both gamma functions in Eq.(\ref{A.2}). If those poles are simple, the Fox function can be expressed in the form of the following series
  \begin{eqnarray} 
\label{A.3}
H_{pq}^{mn}\left[z\left|\begin{array}{c}
(a_p,A_p)\\
\\
(b_q,B_q)
\end{array}\right.\right]=\sum_{h=1}^m\sum_{\nu=0}^\infty\frac{\prod_{j=1,j\ne
h}^m\Gamma(b_j-B_j\frac{b_j+\nu}{B_h})\prod_{j=1}^n\Gamma(1-a_j+A_j\frac{b_h+\nu
}{B_h})}{\prod_{j=m+1}^q\Gamma(1-b_j+B_j\frac{b_j+\nu}{B_h})\prod_{j=n+1}^p
\Gamma(a_j-A_j\frac{b_h+\nu}{B_h})}\frac{(-1)^\nu z^{(b_h+\nu)/B_h}}{\nu!B_h}.
  \end{eqnarray}
A similar expansion can be obtained for $z\to \infty$ by using the property:
  \begin{eqnarray} 
\label{A.4}
H_{pq}^{mn}\left[z\left|\begin{array}{c}
(a_p,A_p)\\
\\
(b_q,B_q)
\end{array}\right.\right]= 
H_{pq}^{mn}\left[\frac{1}{z}\left|\begin{array}{c}
(1-b_q,B_q)\\
\\
(1-a_p,A_p)
\end{array}\right.\right].
  \end{eqnarray}
Another useful property is the multiplication rule:
  \begin{eqnarray} 
\label{A.5}
z^\sigma H_{pq}^{mn}\left[z\left|\begin{array}{c}
(a_p,A_p)\\
\\
(b_q,B_q)
\end{array}\right.\right]= 
H_{pq}^{mn}\left[z\left|\begin{array}{c}
(a_p+\sigma A_p,A_p)\\
\\
(b_q+\sigma B_q,B_q)
\end{array}\right.\right].
  \end{eqnarray}
The Fourier cosine transform of the Fox function yields:
  \begin{eqnarray} 
\label{A.6}
\int_0^\infty H_{pq}^{mn}\left[x\left|\begin{array}{c}
(a_p,A_p)\\
\\
(b_q,B_q)
\end{array}\right.\right]\cos(kx)dx= 
\frac{\pi}{k}H_{q+1,p+2}^{n+1,m}\left[k\left|\begin{array}{l}
(1-b_q,B_q),(1,1/2)\\
\\
(1,1),(1-a_p,A_p),(1,1/2)
\end{array}\right.\right].
  \end{eqnarray}


\bigskip

\begin{thebibliography}{99}

\bibitem{met}
R. Metzler and J. Klafter, Phys. Rep. {\bf 339}, 1 (2000).

\bibitem{bou}
J.-P. Bouchaud and A. Georges, Phys. Rep. {\bf 195}, 12 (1990).

\bibitem{lut}
E. Lutz, Phys. Rev. E {\bf 64}, 051106 (2001).

\bibitem{kar}
C.F.F. Karney, Physica D {\bf 8}, 360 (1983).

\bibitem{zas} 
G.M. Zaslavsky, Phys. Rep. {\bf 371}, 461 (2002).

\bibitem{kubo}
R. Kubo, M. Toda, and N. Hashitsume, {\it Statistical Physics II}
(Springer-Verlag, Berlin, 1985).

\bibitem{bar1}
E. Barkai, Chem. Phys. {\bf 284}, 13 (2002).

\bibitem{sca}
E. Scalas, R. Gorenflo, and F. Mainardi, Phys. Rev. E {\bf 69}, 011107 (2004).

\bibitem{kla1}
J. Klafter and G. Zumofen, J. Phys. Chem. {\bf 98}, 7366 (1994).

\bibitem{bar}
E. Barkai, R. Metzler, and J. Klafter, Phys. Rev. E {\bf 61}, 132 (2000).

\bibitem{osh}
B. O'Shaughnessy and I. Procaccia, Phys. Rev. Lett. {\bf 54}, 455 (1985).

\bibitem{kwo}
Kwok Sau Fa, Phys. Rev. E {\bf 72}, 020101(R) (2005).

\bibitem{com}
A. Compte, Phys. Rev. E {\bf 53}, 4191 (1996).

\bibitem{met1}
R. Metzler and T. F. Nonnenmacher, Chem. Phys. {\bf 284}, 67 (2002).

\bibitem{met2}
R. Metzler, J. Klafter, and I. Sokolov, Phys. Rev. E {\bf 58}, 1621 (1998).

\bibitem{com1}
A. Compte, Phys. Rev. E {\bf 55}, 6821 (1997).

\bibitem{fog}
H. C. Fogedby, Phys. Rev. Lett. {\bf 73}, 2517 (1994).

\bibitem{com2}
A. Compte and M. O. C\'aceres, Phys. Rev. Lett. {\bf 81}, 3140 (1998).

\bibitem{kam1}
A. Kami\'nska and T. Srokowski, Phys. Rev. E  {\bf 69}, 062103 (2004).

\bibitem{fri}
A. Brissaud and U. Frisch, J. Math. Phys. {\bf 15}, 524 (1974).

\bibitem{kam}
A.Kami\'nska and T. Srokowski, Phys. Rev. E {\bf 67}, 061114 (2003).

\bibitem{san}
R. S\'anchez, B. A. Carreras, and B. Ph. van Milligen, 
Phys. Rev. E {\bf 71}, 011111 (2005).

\bibitem{sro}
T. Srokowski and A. Kami\'nska, Phys. Rev. E {\bf 70}, 051102 (2004).

\bibitem{gar}
C. W. Gardiner, {\it Handbook of Stochastic Methods for Physics, Chemistry
and the Natural Sciences} (Springer-Verlag, Berlin, 1985).

\bibitem{vkam}
N. G. van Kampen, {\it Stochastic Processes in Physics and Chemistry}
(North-Holland, Amsterdam, 1981).

\bibitem{ved}
A. A. Vedenov, Rev. Plasma Phys. {\bf 3}, 229 (1967).

\bibitem{fuj}
H. Fujisaka, S. Grossmann, and S. Thomae, 
Z. Naturforsch. Teil A {\bf 40}, 867 (1985).

\bibitem{kwo1}
Kwok Sau Fa and E. K. Lenzi, Phys. Rev. E {\bf 67}, 061105 (2003).

\bibitem{hen}
H. G. E. Hentschel and Procaccia, Phys. Rev. A {\bf 29}, 1461 (1984).

\bibitem{old}
K. B. Oldham and J. Spanier, {\it The Fractional Calculus}, (Academic Press, San Diego, 1974).

\bibitem{sch}
W. R. Schneider, in: S. Albeverio, G. Casati, D. Merlini (Eds.), Stochastic Processes in Classical and Quantum Systems, Lecture Notes in Physics, Vol. 262, Springer, Berlin, 1986.

\bibitem{wes}
B. J. West, P. Grigolini, R. Metzler, and T. F. Nonnenmacher,
Phys. Rev. E {\bf 55}, 99 (1997).

\bibitem{cha}
J. M. Chambers, C. L. Mallows, and B. W. Stuck,
J. Amer. Statist. Assoc. {\bf 82}, 704 (1987).

\bibitem{fox}
C. Fox, Trans. Am. Math. Soc. {\bf 98}, 395 (1961).

\bibitem{mat}
A. M. Mathai and R. K. Saxena, The $H$-function with Applications in Statistics and Other Disciplines, Wiley Eastern Ltd., New Delhi, 1978.

\bibitem{sri}
H. M. Srivastava, K. C. Gupta, and S. P. Goyal, The $H$-functions of one and two variables with applications, South Asian Publishers, New Delhi, 1982.


\end{thebibliography}
\end{document}